\documentclass[prb,twocolumn]{revtex4}

\usepackage{graphicx}

\begin{document}

\title{Dynamics of stripe patterns in type-I superconductors subject to a rotating field}

\author{M.~Menghini and R.~J.~Wijngaarden}
\affiliation{Department of Physics and Astronomy, Vrije
Universiteit, De Boelelaan 1081, 1081HV Amsterdam, The
Netherlands}

\date{ \today}

\begin{abstract}

The evolution of stripe patterns in type-I superconductors subject
to a rotating in-plane magnetic field is investigated
magneto-optically. The experimental results reveal a very rich and
interesting behavior of the patterns. For small rotation angles, a
small parallel displacement of the main part of the stripes and a
co-rotation of their very ends is observed. For larger angles,
small sideward protrusions develop, which then generate a zigzag
instability, ultimately leading to a breaking of stripes into
smaller segments. The short segments then start to co-rotate with
the applied field although they lag behind by approximately
$10^\circ$.  Very interestingly, if the rotation is continued,
also reconnection of segments into longer stripes takes place.
These observations demonstrate the importance of pinning in type-I
superconductors.

\end{abstract}


\maketitle

Type-I superconductors have been studied quite extensively since
their discovery and until the 60s. Even though the formation of
patterns in the intermediate state  was observed  by different
techniques \cite{huebener}, little attention was paid to the {\it
dynamic} properties of this modulated phase. In this work we focus
on the experimental study of the dynamics of stripe patterns in
the intermediate state of type-I superconductors.  A thin sheet of
type-I material in a magnetic field perpendicular to its large
surface is in the IS for a certain range of temperatures and
magnetic fields. In this state, alternating normal (N) ($B \neq
0$) and superconducting (SC) (B = 0) macroscopic regions are
formed in the sample. This modulated phase is a consequence of the
competition between the magnetic energy that favors the formation
of small domains and a positive SC-N interface energy that tends
to form large domains. It is well known \cite{sharvin} that
straight laminae of SC and N domains are formed when the sample is
cooled in a magnetic field with non-zero in-plane component,
$H_{xy}$ (Sharvin geometry). The laminae are extended along the
direction of $H_{xy}$, with period and width determined by the
applied field and the sample thickness. R. Goldstein {\it et al.}
\cite{goldstein2,goldstein1} developed the theoretical {\it
current loop model} to explain the formation of patterns in type-I
superconductors. They consider the IS as a collection of current
ribbons flowing along the SC-N interfaces. Even though the
magnetic energy is treated only approximately \cite{narayan} the
model reproduces the results of the Landau model for straight
laminae \cite{landau}. Instabilities of the laminar state in
type-I superconductors are expected \cite{goldstein2} similarly to
other systems. For example, an Eckhaus instability \cite{eckhaus}
should appear when the field changes slowly due to rearrangement
of the laminae width and period to the new equilibrium condition.
In convective systems, \cite{croquette,lowe} this instability
leads to a smooth movement of dislocations. In addition, it was
predicted \cite{goldstein2} that a zigzag pattern is induced in
the IS of a type-I superconductor if $H_{xy}$ is rotated with
respect to the direction of the laminae.

\begin{figure}[hbb]
\centering
\includegraphics[width=80mm,angle=0]{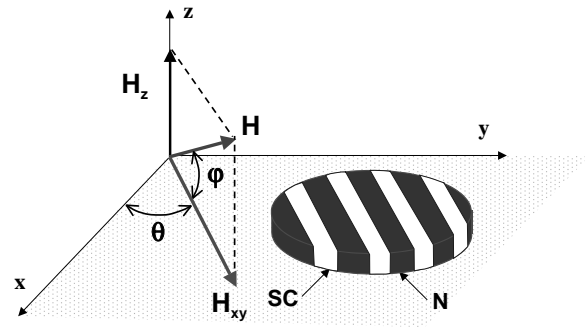}
\caption{Schematic of sample and  magnetic field configuration.
$H$ is the applied magnetic field vector, $H_{xy}$ is the
projection of $H$ onto the plane of the superconductor.}
  \label{sample}
\end{figure}

\begin{figure}[ttt]
\centering
\includegraphics[width=80mm,angle=0]{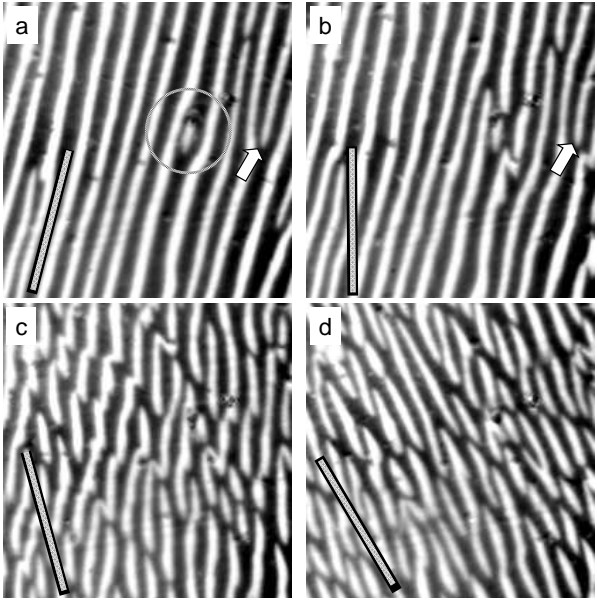}
\caption{Sequence of images for different directions of the
applied in-plane field, $H_{xy}$. (a)$\theta=0^\circ$,
(b)$\theta=15^\circ$, (c)$\theta=30^\circ$,(d)$\theta=45^\circ$.
The straight lines on the lower left  indicate the direction of
$H_{xy}$. White (black) corresponds to superconducting (normal)
regions and the scale bar corresponds to 0.5 mm.}
  \label{images}
\end{figure}

We study the response of a stripe pattern in a type-I
superconductor to a change of in-plane field direction. The sample
was obtained by flattening a Pb foil down to a thickness of 0.15
mm and by  cutting it to  a lateral dimension of $\sim 1$ cm. The
sample was annealed at $250 ^\circ$C to remove stress induced by
flattening and cutting. We present experiments made in a square
sample but equivalent results were obtained in a circular shaped
one. Magneto-optical images were acquired using the technique
described in \cite{wijngaarden} which yields maps of the local
magnetic induction $B$ just above the surface of the sample. The
sample is mounted in a Oxford Instruments 7T vector magnet system
that allows to change independently 3 perpendicular components of
the magnetic field. We define $\theta$ as the angle between the
projection of the field in the xy plane ($H_{xy}$) and the x-axis
and $\varphi$ as the angle between the total field vector and
$H_{xy}$, see Fig.\,\ref{sample}. The experiments are performed as
follows. First, an ordered stripe structure is obtained at 6 K by
decreasing $H_{x}$ from large values such that the sample is in
the normal phase, down to 26 mT while $H_z$ is fixed at 2 mT.
These values correspond to $\varphi=4.4^\circ$ and
$\theta=0^\circ$. Subsequently, $\theta$ is changed in steps of
$1^{\circ}$ at a rate of $0.12^\circ/$sec while $\varphi$ is kept
constant. A typical initial pattern  is shown in
Fig.\,\ref{images}(a). Similar structures are obtained if the IS
is reached decreasing $H_z$ at constant $H_{xy}$ or by decreasing
temperature at constant $H$.

In the initial state ($\theta=0^\circ$, Fig. \ref{images}(a)) some
 dislocations in the form of stripes that end in the interior
of the sample are observed. This is a consequence of a stripe
period re-adjustment when decreasing the (in-plane) field. The
energy of a dislocation is found to be  very small
\cite{goldstein2} indicating that, as we observe experimentally,
it is easy to nucleate this type of defects in the IS of type-I
superconductors. As in the case of convective systems
\cite{croquette, lowe}, the formation of edge dislocations is a
consequence of an Eckhaus instability.

Up to $\theta=15^\circ$ (Fig.\,\ref{images}(b)) the pattern
remains quite ordered and the stripes  are mostly parallel to the
initial direction ($\theta=0^{\circ}$). It is also evident that
the rotation has induced the breaking of a few stripes. New
free-ends of stripes are formed in regions where the laminae are
already distorted in the initial state, see the encircled region
in Fig. \ref{images}(a). The free-ends are oriented along $H_{xy}$
while the rest of the stripe remains locked at $\theta= 0^\circ$.
This suggests that the extra surface energy when breaking the
stripe is compensated by the energetic gain of orienting it
parallel to $H_{xy}$. Consistently, some of the already present
free-ends in the initial pattern (see arrows in Fig.
\ref{images}(a) and (b)) also align with the in-plane field for
$\theta= 15^\circ$. Also the {\it ends} of the stripes at the edge
of the sample (not shown) show a tendency to follow the actual
field direction, while the remaining portion of the stripes are
locked at $\theta=0^\circ$.

\begin{figure}[ttt]
\centering
\includegraphics[width=80mm]{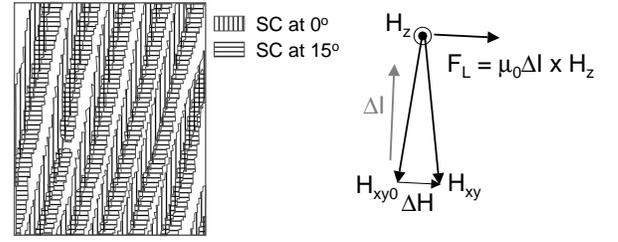}
\caption{Left panel:  overlap of the positions of the SC domains
at $\theta = 0^\circ$ and $15 ^\circ$ showing the displacement of
stripes towards the right of the image. Right panel: schematic
drawing of the Lorentz force $F_{L}$ on the top surface of the
sample, induced by the change in $H_{xy}$ direction.}
  \label{subt}
\end{figure}

Besides the disorder  induced by the reorientation of stripe-ends
when rotating over $15^\circ$, a small net displacement of the
structure is also observed, see left panel of Fig\,\ref{subt}.
Vertical-dashed (horizontal-dashed)  regions indicate the stripes
in the structure at $\theta=0^\circ$ ($\theta=15^\circ$). A change
in the direction of the in-plane field component induces currents
that flow on the top and bottom surfaces of the sample. The
Lorentz force associated with these screening currents causes the
translation of stripes as observed by MO experiments on the top
surface (Fig. \ref{subt}, right panel). On the bottom surface the
force has opposite sign (since the current flows in opposite
direction), leading to a shear deformation of the stripes. Thus,
the main effect of the rotation in this first stage is a shear of
the laminae along the sample thickness detected as a translation
motion on the top surface. This induced Lorentz (shear) force is
acting on the stripes during the whole rotation experiment.
However, its effect is more evident at the beginning. At a later
stage, the dynamics of the stripes is rather complex making the
observation of the translation motion more difficult.

If the angle of rotation increases to $30^{\circ}$  (Fig.
\ref{images}(c)), in some regions  zigzag-like structures are
formed in agreement with theoretical prediction \cite{goldstein2}.
Surprisingly, cutting of some long stripes is also observed giving
rise to the formation of short SC segments. Finally, at
$\theta=45^{\circ}$ (Fig. \ref{images}(d)) the pattern consists of
short segments, either straight or s-shaped. It is interesting to
note that if $H_{xy}$ is turned back to $\theta=0^\circ$ is not
possible to recover the same pattern as at the beginning of the
experiment. Instead, a more disordered structure remains that has
higher energy than the originally ordered one. If the rotation is
stopped at a certain angle and the external parameters are kept
constant no significant changes in the structure are observed over
a long time ($ \sim 1$ hour). These two results indicate that
after the rotation, the system is in a local minimum of energy
with very large relaxation time constant. This implies significant
energy barriers induced by the pinning potential in the sample.

\begin{figure}[ttt]
\centering
\includegraphics[width=80mm]{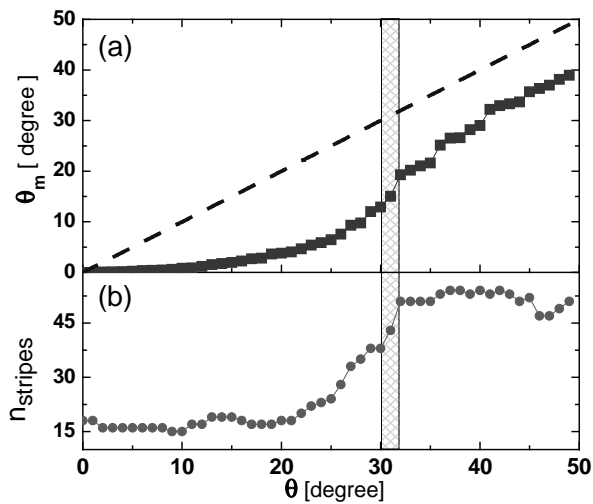}
\caption{(a) Average angle of the pattern, $\theta_m$, (square
symbols) as a function of the angle of the applied magnetic field,
$\theta$. The dashed line corresponds to $\theta_m=\theta$. (b)
Number of stripes as a function of $\theta$. Note the abrupt
change at $\theta=30^\circ$, where most of the laminae break into
shorter segments.}
  \label{angle}
\end{figure}

In order to investigate how the patterns  follow the in-plane
field  direction we calculate  the average direction of the
pattern, $\theta_{m}$, from the average orientation of the peaks
in the Fourier transform of the images. The results of this
calculation are shown in Fig.\,\ref{angle}(a). The dashed line
corresponds to the condition that the pattern is parallel to
$H_{xy}$ and the square symbols are the experimental values. As we
have mentioned, at small angles $\theta \lesssim 13^\circ$, the
orientation of the structure is locked at the direction defined by
the initial $H_{xy}$. At larger angles, the mean direction changes
as the stripes try to follow the in-plane field and at $\theta
\sim 30^\circ$  a rapid increase is observed. In order to find
 the origin of this jump in $\theta_m$ we calculate the
number of segments, $n_{stripes}$, as a function of the angle of
rotation (see Fig. \,\ref{angle}(b)). Clearly,  at $\theta =
30^\circ$ there is also a rapid increase of $n_{stripes}$
indicating that the improved alignment of the structure is related
with the breaking of many stripes. At $\theta= 35^\circ$ and
$40^\circ$ also small jumps in $\theta_m$ are observed. However,
in those cases there is no significant change in $n_{stripes}$.
Here the realignment is made by rotation of the already short
segments. Thus, we find that the mechanism for alignment of the
stripes with the direction of the in-plane field is: cutting of
stripes followed by rotation of short segments.

From the present experimental results it is clear that pinning
plays an important role in the dynamics of these structures. The
stripes remain locked at the initial direction, even when the
direction of $H_{xy}$ is changed by $\sim 13^\circ$. This is not
due to the stripes being locked by extended defects, because when
$\theta=360^\circ$ is reached and the rotation continues
thereafter, no lock-in of the structure is observed. In addition,
when the experiment is repeated under the same conditions, the
stripes do not nucleate in exactly the same positions, nor do they
return to the same positions when the field is first rotated to
say $\theta \sim 45^\circ$ and subsequently to $\theta \sim
0^\circ$. In the last case, always a kind of "backlog" is
observed. We conclude that pinning is important and that it is the
result of many small defects acting together. When the structure
is nucleated, the system takes as much profit as possible from the
pinning sites. As the angle of rotation increases there is an
increase in the magnetic energy due to misalignment with the
external field. At a certain moment, the stripes start to break
and rotate, while being still partially pinned: now they are in a
metastable state and do not profit fully anymore from the energy
lowering by pinning.

\begin{figure}[ttt]
\centering
\includegraphics[width=80mm,angle=0]{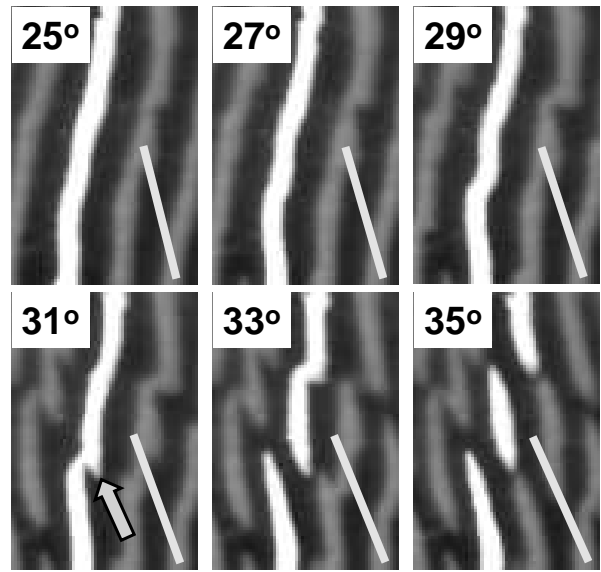}
\caption{Zoomed-in images of a rotation experiment showing the
zigzag instability and the cutting of stripes. The original image
is manipulated in order to highlight one particular stripe as the
angle is changed. The straight lines indicate the direction of the
applied field. } \label{detailcut}
\end{figure}

In  Fig. \ref{detailcut} we show the evolution of one SC stripe
(highlighted) as the direction of $H_{xy}$ changes such that the
process of zigzag formation and subsequent cutting of stripes can
be observed. The straight lines indicate the direction of
$H_{xy}$. At $\theta=25^\circ$ the segment is misaligned with
respect to $H_{xy}$ by $20^\circ$. As $\theta$ increases to
$27^\circ$  the segment bends. Then, the undulations of the stripe
become sharper giving rise to the onset of  a zigzag structure
($\theta = 29^\circ$). At $\theta = 33^\circ$ the stripe breaks in
the lower part while at approximately 1/3 of its original length a
clear zigzag is observed. Finally, at $\theta=35^\circ$ the line
breaks also at that point and the lowest and highest segments are
reasonably aligned with the actual $H_{xy}$ while the middle
segment is misaligned by about $10^\circ$. In subsequent images
(not-shown) the middle segment also rotates in response to the
change in $\theta$. Apparently, a zigzag is an intermediate
structure formed prior to the cutting of stripes. The zigzags seem
to be initiated by small sideward protrusions to the stripes (see
arrow in the image at $\theta = 31^\circ$), which have roughly the
same period (measured in the direction normal to the actual
$H_{xy}$) as the original stripes. These protrusions can be seen
as new laminae growing parallel to the actual $H_{xy}$ from the
existing laminae. However, before the protrusions become very
large, the original laminae break and their ends turn in the
direction of the actual field, thereby absorbing the protrusions.
As soon as the laminae break, the applied field generates a torque
on the new-free ends promoting their rotation and consequent
alignment with $H_{xy}$ (see the image at $\theta= 33^\circ$ in
Fig. \ref{detailcut}). This torque also induces s-shaped deformed
segments. Similar effects have been observed in chains of magnetic
nanoparticles subjected to a rotating magnetic field \cite{melle}.

\begin{figure}[ttt]
\centering
\includegraphics[width=80mm,angle=0]{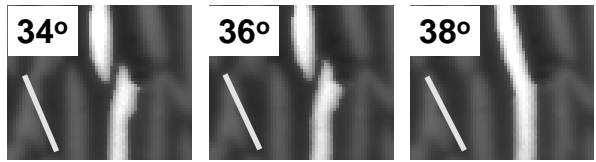}
\caption{Same as Figure \ref{detailcut} showing a reconnection
event.}
  \label{reconnect}
\end{figure}

The inhomogeneous distribution of magnetic field inside the sample
implies the presence of currents at the SC-N interface (that
shield the magnetic field from the SC regions). Hence, the cutting
of stripes  implies that neighboring regions (at the new free-ends
of stripes) with currents flowing in {\it opposite} directions are
formed.  Considering only interaction between currents, the
stripe-ends repel each other. Thus, in this simple picture,
reconnection of stripes is unexpected. However, we do  observe
reconnection of SC stripes, see Fig. \ref{reconnect}. This
suggests that the forces associated with the interface energy and
magnetic energy are able to overcome the repulsion between the
currents. We have also observed that the locations in the sample
where the cutting and reconnection takes place are not
reproducible. Therefore we can rule out that these events are
induced by irregularities or static defects in the sample.

Very recently, numerical simulations of the response of stripes in
a {\it magnetic} system subjected to a rotating in-plane field
were performed \cite{jaglaprivate}. The preliminary results show
that the stripe domains first cut and later partially reconnect
while the field is rotating, as observed in our experiments. Since
the same model of simulations reproduces various patterns formed
in type-I superconductors \cite{jaglapre} it is reasonable to
think that the cutting and reconnection of  stripes is an
intrinsic behavior of a stripe pattern subjected to a rotating
field.

The results presented in this article show that the evolution of a
stripe pattern subject to a rotating field is more complex and
rich than expected. At small rotation angles the whole structure
(except at the edges of the sample) is locked at the direction
fixed by the initial condition. In this regime, there is only a
shear of the stripes along the thickness of the sample due to the
currents induced by the change in $H_{xy}$ direction. The
corresponding Lorentz force is due to surface currents and hence
acts mainly on those parts of the stripes that are in contact with
the sample surface. The bulk of the stripe is, however, free from
this Lorentz force. As a consequence, a small shear of the stripe
can happen rather easily (large Lorentz force/pinning force
ratio), while rotation of a complete stripe involves the bulk
pinning force on the whole stripe and is much  more unfavorable
(small Lorentz force/pinning force ratio). Hence, we observe first
a shift of the stripes and only at later stage the rotation
starts.  Subsequently, a bending of stripes and a zigzag
instability, in agreement with the theoretical expectation
\cite{goldstein2}, is observed. As the angle of rotation
increases, cutting and reconnection of stripes is detected. The
rotation of the stripes is the result of the balance between the
torque due to the misalignment of  the stripes with respect to the
applied field and the pinning force. Our experimental results
indicate that a delicate balance between all the terms in the
energy determines the response of a stripe structure to a rotating
field. In particular, it is evident that pinning plays an
important role. This brings up a new challenge in the
understanding of the dynamics of modulated phases and a more
stringent test for current string models for type-I
superconductors.


 \acknowledgments
 This work was supported by FOM (Stichting voor Fundamenteel Onderzoek der
Materie) which is financially supported by NWO (Nederlandse
Organisatie voor  Wetenschappenlijk Onderzoek).


\end{document}